\documentstyle[12pt]{article}

\begin{document}

\title{Ricci Collineations \\
for \\
type B warped space-times}
\author{J.Carot\thanks{%
Email: dfsjcg0@ps.uib.es} \\
Departament de F\'{\i}sica\\
Universitat de les Illes Balears\\
E-07071 Palma de Mallorca\\
Spain \and L. A. N\'{u}\~{n}ez \thanks{%
Email:nunez@ciens.ula.ve} and U. Percoco \thanks{%
Email: upercoco@ciens.ula.ve} \\
{\it Centro de Astrof\'\i sica Te\'orica}\\
{\it Departamento de F\'\i sica, Facultad de Ciencias, }\\
Universidad de los Andes, M\'{e}rida 5101, Venezuela.}
\maketitle

\begin{abstract}
We present the general structure of proper Ricci Collineations (RC) for type
B warped space-times. Within this framework, we give a detailed description
of the most general proper RC for spherically symmetric metrics. As
examples, static spherically symmetric and Friedmann-Robertson-Walker
space-times are considered.
\end{abstract}

\section{Introduction}

The purpose of this paper is to study Ricci Collineations (RC) for a certain
class of space-times, namely type B warped space-times and in particular
spherically symmetric space-times. Collineations are symmetry properties of
space times. Katzin{\it \ et al. }\cite{KatzinEtal69} define them as those
vector fields, $X$, such that leave the various relevant geometric
quantities in General Relativity invariant under Lie dragging in their
direction. The best known examples of collineations are the {\it Killing
vectors} ({\it Motions}), i.e. vectors that satisfy: 
\begin{equation}
\pounds _X\ g_{ab}=0  \label{collkill}
\end{equation}
Other interesting symmetries are defined in analogously and the more
frecuent cases of study have been:\\Conformal Motions: 
\begin{equation}
\pounds _X\ g_{ab}=2\sigma g_{ab}  \label{collconf}
\end{equation}
Affine Collineations: 
\begin{equation}
\pounds _X\ \Gamma _{ab}^c=0  \label{collafin}
\end{equation}
Curvature Collineations: 
\begin{equation}
\pounds _X\ R_{abcd}=0  \label{collcurv}
\end{equation}
Ricci Collineations: 
\begin{equation}
\pounds _X\ R_{ab}=0  \label{collricc}
\end{equation}
Contracted Ricci Collineations: 
\begin{equation}
g^{ab}\pounds _X\ R_{ab}=0  \label{collcric}
\end{equation}
Here $\pounds _X$ stands for the Lie derivative operator and the indices $%
a,b,...$ run from 1 to 4.

The well established connection between {\it Killing Vectors } and constants
of the motion has encouraged the search for general relations between
collineations and conservation laws. Collineations, other than {\it Motions}%
, can be considered as non-noetherian symmetries and can also be associated
to constants of the motion. {\it Affine Collineations} have been shown to be
related to conserved quantities \cite{HojmanEtal86}, and this property has
been used to integrate geodesics of the Robertson-Walker metric \cite
{BedranLesche86}. As far as we know, the first {\it Curvature Collineation}
was found by Aichelburg \cite{Aichelburg70} for pp-wave metrics, and their
relationships to first integrals of the geodesic equations extensively
studied in Ref. \cite{KatzinEtal69}. Particular types of {\it Ricci} and 
{\it Contracted Ricci Collineations}, for the Robertson-Walker metric have
also been found and shown to be related to the particle number conservation 
\cite{GreenEtal77}. Also, considerable attention is being paid to the
related problem of symmetry inheritance in General Relativity \cite
{ColeyTupper89}. Collineations have been studied in connection with fluid
spacetimes \cite{GreenEtal77}, \cite{OliverDavis77}, \cite{TsamparlisMason90}%
, \cite{Duggal92} and some specific examples have been given for the {\it %
C-metric } \cite{AulestiaEtal84}, Robertson-Walker Spacetimes \cite
{NunezEtal90}, and G\"{o}del-type manifolds \cite{MelfoEtal92}.

It is clear from the above definitions that {\it Motions} are particular
cases of {\it Affine Collineations}, {\it Affine Collineations} are
particular cases of {\it Curvature Collineations}, and so on. It is
therefore possible to construct an ``inclusion diagram'' connecting these
symmetries. One such diagram, that includes these and other related
symmetries, is presented in Ref. \cite{KatzinEtal69}. A collineation of a
given type is said to be{\it \ proper} if it does not belong to any of the
subtypes. Clearly, in solving any collination equation, with the obvious
exception of the {\it Killing equation}, solutions representing improper
collineations can be found. Therefore, in order to relate a symmetry to a
particular conservation law, and its corresponding constant of the motion,
the ``properness'' of that collineation must be assured. Some computer
algebra tools have been developed to check the properness of Ricci and other
collineation vectors are under development \cite{MelfoNunez92} \cite
{BertolotiEtal95a}.

We assume that RCs are smooth vector fields. Although this is not
necessarily so, by restricting ourselves to this case, we ensure that they
form a Lie algebra with the usual bracket operation. Such an algebra
naturally contains that of Special Conformal Killing Vectors (SCKV) (see
reference \cite{ColeyTupper89}) which in turn contains that of Homothetic
Vector Fields (HVF) and therefore the isometry algebra of all Killing Vector
Fields (KV).

Regarding the Ricci tensor, we shall consider that it is non-degenerate
(i.e.: rank 4) and this in turn ensures that the Lie algebra of RC is finite
dimensional, its maximal dimension being 10 (9 being forbidden by Fubini 's
theorem). For further information on issues concerning dimensionality and
degenerate Ricci tensor see, for instance, references \cite{CarotEtal94} and 
\cite{HallEtal95}.

The paper is organized as follows: in section 2 we describe the basic
features of the RC in type B warped space-times, then, in section 3, we
consider spherically symmetric space-times as a particular case of these,
studying two distinct cases, namely: static solutions and
Friedmann-Robertson-Walker space-times.

\section{Type B warped space-times}

Suppose that $(M_1,h_1)$ and $(M_2,h_2)$ are a pair of pseudo-Riemannian
manifolds, and $\Phi$ is a real valued function on $M_1$ (\lq warping
function'), one can then build a Lorentz manifold, $(M,g)$ by setting $M=M_1
\times M_2$ and $g=\pi_1^\ast h_1 \otimes \Phi^2 \pi_2^\ast h_2$, where the
functions $\pi_1$ and $\pi_2$ are the canonical projections onto the factors
of the product. $(M,g)$ is then called a \lq warped product manifold'. If
dim $M=4$, we say that $(M,g)$ is a \lq warped space-time' and one can
classify them according to the respective dimensions of the factor
(sub-)manifolds $M_1$ and $M_2$. We shall refer the reader to \cite
{CarotDaCosta93} and references cited therein for a general discussion,
restricting ourselves hereafter to the case dim $M_1=$ dim $M_2=2$, namely;
warped space-times of the class $B$. Although all our considerations will be
local, see \cite{Haddow} for some remarks on globally warped space-times. It
can be shown that for type $B$ warped space-times, a coordinate chart exists
(adapted to the manifold product structure), such that the metric takes the
form

\begin{equation}  \label{warped}
ds^2=h_{AB}\left( x^D\right) \ {\rm d}x^A\ {\rm d}x^B+ \Phi ^2\left(
x^D\right) \ h_{\alpha \beta }\left( x^\gamma \right) \ {\rm d}x^\alpha \ 
{\rm d}x^\beta
\end{equation}
where the indices $A,B,\ldots $ run from 1 to 2 and $\alpha ,\beta , \ldots $
from 3 to 4. The functions $h_{AB}$ and $h_{\alpha \beta }$ are the
component forms of $\pi_1^\ast h_1$ and $\pi_2^\ast h_2$ in the local charts 
$\{ x^A \}$ and $\{ x^\alpha \}$, which are in turn adapted to $M_1$ and $%
M_2 $ respectively. The Ricci tensor of such a space-time takes then the
following component form in the above chart:

\begin{equation}  \label{ricci1}
R_{AB}=\frac 12R_1\ h_{AB}-\frac 2\Phi \ \Phi _{A;B}\ ,
\end{equation}
\begin{equation}  \label{ricci2}
R_{A\alpha }=0\ ,
\end{equation}
\begin{equation}  \label{ricci3}
R_{\alpha \beta }=\frac 12\left( R_2-\left( \Phi ^2\right) _{;A}^A\right)
h_{\alpha \beta }\ \equiv \ Fh_{\alpha \beta } \ ;
\end{equation}
where $F=\frac 12\left( R_2-\left( \Phi ^2\right) _{;A}^A\right)$ and $R_1$
and $R_2$ are the Ricci scalars associated to the 2-metrics $h_1$ and $h_2$.
The semi-colon indicates, as usual, the covariant derivative with respect to
the space-time metric.

Let now $X$ be a RC on $M$, and define its vertical and horizontal
components, $X_1$ and $X_2$, as follows (see \cite{CarotDaCosta93}):

\begin{equation}
X_1^a\equiv g^{ab}\left( \pi _1^{*}h_1\right) _{bd}X^d\ \ \ \ X_2^a\equiv
X^a-X_1^a  \label{components}
\end{equation}
In the above adapted chart, one readily sees that $X_1^A=X^A,$ $\ X_1^\alpha
=0\ $, and $X_2^A=0,\ X_2^\alpha =X^\alpha \ $.

On account of (\ref{ricci1}), (\ref{ricci2}) , ( \ref{ricci3}) and (\ref
{components}), equation (\ref{collricc}) is now equivalent to:

\begin{equation}  \label{set}
R_{AB,D}X_1^D + R_{AD}X^D_{1,B} + R_{DB}X^D_{1,A}=0 \ ,
\end{equation}

\begin{equation}  \label{vuit}
R_{AD}X^D_{1,\alpha} + Fh_{\alpha \beta}X^\beta_{2,A}=0 \ ,
\end{equation}

\begin{equation}
\pounds _{X_2}h_{\alpha \beta }=2\Psi h_{\alpha \beta }  \label{nou}
\end{equation}
where 
\begin{equation}
\Psi =-\frac 12{\frac{F_{,D}X_1^D+F_{,\gamma }X_2^\gamma }F}  \label{deu}
\end{equation}
Take now $p_1\in M_1$ and consider the manifold ${\tilde{M}}_2\equiv
\{p_1\}\times M_2\cong M_2$ (see \cite{CarotDaCosta93}), equation (\ref{nou}%
) is then a statement that $X_2$ is a Conformal Killing Vector (CKV) of $({%
\tilde{M}}_2,h_2)$, and therefore it can be re-written as 
\begin{equation}
X_{2\alpha /\beta }+X_{2\beta /\alpha }=2\Psi \ h_{\alpha \beta }
\label{x2conf1}
\end{equation}
where a stroke denotes the covariant derivative associated with the metric $%
h_2$.

Furthermore, it is possible to write \cite{Hall90}, 
\begin{equation}
\pounds _{X_2}R_{2\alpha \beta }=-2\Psi _{\alpha /\beta }-\left( h^{\mu \nu
}\Psi _{\mu /\nu }\right) h_{\alpha \beta }  \label{x2conf2}
\end{equation}
where 
\begin{equation}
R_{2\alpha \beta }~=~\frac{R_2}2\ h_{\alpha \beta }  \label{Escricci2}
\end{equation}
is the Ricci tensor of the metric $h_2$ . In addition, the Conformal
Bivector associated to $X_2$ , i.e. 
\begin{equation}
F_{\alpha \beta }\equiv X_{2\alpha /\beta }-X_{2\beta /\alpha }
\label{Bivector}
\end{equation}
satisfies 
\begin{equation}
F_{\alpha \beta /\gamma }=\frac{R_2}2(h_{\alpha \gamma }\ X_{2\beta
}-X_{2\alpha }\ h_{\beta \gamma })-\Psi _\alpha \ h_{\beta \gamma }+\Psi
_\beta \ h_{\alpha \gamma }  \label{x2conf3}
\end{equation}
Now, from (\ref{x2conf2}) one obtains 
\begin{equation}
\Psi _{\alpha /\beta }=\lambda h_{\alpha \beta }\qquad {\rm and}\qquad
\lambda \equiv -\frac 18(\pounds _{X_2}\ R_2+2\Psi \ R_2)  \label{x2conf4}
\end{equation}
and from the Bianchi identities (on $({\tilde{M}}_2,h_2)$) for $\Psi \ $, it
readily follows: 
\begin{equation}
\lambda _{,\gamma }=-\frac{R_2}2\Psi _{,\gamma }  \label{lambdaderiv}
\end{equation}
Furthermore, taking a further covariant derivative in the above expression,
skewsymmetrising, and equating to zero, one has 
\begin{equation}
-\left( \frac{R_2}2\right) _{,\alpha }=\sigma \Psi _{,\alpha }
\label{R2deriv}
\end{equation}
for some function $\sigma $.

To proceed with our study, it is useful to consider now the following
decomposition of ${\tilde M}_2$; ${\tilde M}_2= H \cup K \cup C$, where $H$
is that open submanifold of ${\tilde M}_2$ on which $\Psi_{\alpha / \beta} 
\not{= }0$ (hence $\lambda \not{= }0$ and $\Psi^\alpha \Psi_\alpha \not{=}0$
on $H$), $K$ is the interior of that set of points for which $\Psi_{\alpha /
\beta} = 0 \ $, and $C$ is a set with no interior defined by the
decomposition itself.

We shall first study what happens in $K$. Since $\Psi_{\alpha / \beta} = 0 $
there, it follows that $\Psi_{, \alpha}$ is either zero on $K$ (in which
case $X_2$ is homothetic), or else it is a (gradient) Killing vector (and
then $X_2$ is an SCKV), the Bianchi identities then implying $R_2=0 \ $,
i.e.; $h_2 \vert_K$ is flat. In the latter case ($h_2$ flat), one can always
choose coordinates on $K$, say $\{ x,y \} \ $, such that $\Psi \vert_K = A x$
($A =$ constant), and integrating out the conformal equations (\ref{nou})
for $X_2$ on $K$ it follows

\begin{equation}  \label{sckvK}
X_2= \left( \frac 12 A(x^2-y^2) - D y + L \right) \partial_x + \left( Axy +
Dx + E \right) \partial_y
\end{equation}
where $A, \ D, \ E$ and $L$ are constants on $K$ which will, in general,
depend on the chosen $p_1 \in M_1$, and therefore, when considering $X_2$ on 
$M$, one will have that all of them are functions of the coordinates set up
in $M_1$, thus $A=A(x^B), \ D=D(x^B), \cdots$ to be determined, along with
the vertical component $X_1$ of $X$, from (\ref{set}) and (\ref{vuit}). In
fact, it is easy to see from (\ref{vuit}) that $A$ and $D$ must be
constants, say $A=A_0$ and $D=D_0$, and from the expression (\ref{deu}) of $%
\Psi$ with $R_2=0 \ $, together with $\Psi=A_0x $, it follows that $E$ must
also be constant (which can be set equal to zero without loss of
generality), then from (\ref{set}) $X_1^A=P^A(x^B)x+Q^A(x^B)$, and therefore
one has, on $M \cap K$ and if $R_2=0$: 
\begin{equation}  \label{2flat1}
X= ( P^Ax+Q^A ) \partial_A + ( \frac {A_0}2 (x^2-y^2) - D_0 y + L)
\partial_x + ( A_0 xy + D_0 x ) \partial_y
\end{equation}
\begin{equation}  \label{2flat2}
\Psi=A_0x
\end{equation}
$P$, $Q$ and $L$ being functions of the coordinates $\{ x^B \} $ on $M_1$ to
be determined from (\ref{set}) and (\ref{vuit}).

If $\Psi _{,\alpha }|_K=0$, $X_2$ is an HVF, and therefore (\cite{Hall90}) $%
\pounds _{X_2}R_2=-2\Psi R_2$ if $R_2\neq $ constant or $\pounds _{X_2}R_2=0$
if $R_2=$ constant, hence (\ref{deu}) implies 
\begin{equation}
\Psi =-\frac 12{\frac{(\Phi ^2)_{;AD}^AX_1^D}{(\Phi ^2)_{;A}^A}}
\label{psinova}
\end{equation}
thus, given a basis of the homothetic algebra of $(M_2,h_2)$, say $\{\zeta
_I\}$ with $I\leq 4$, one will have $X_2=C^I\zeta _I$ on $({\tilde{M}}%
_2,h_2) $, the $C^I$ being constants which will in general depend on the
chosen $p_1\in M_1$, and again, when considering $X_2$ on $M$, they will
become functions of the coordinates in $M_1$, to be determined as before
from (\ref{set}) and (\ref{vuit}). It is worth noticing that, whenever a
proper HVF exists in $({\tilde{M}}_2,h_2)$, say $\zeta _1$ then (\ref{nou})
implies that $C^1=\Psi $. It will be shown later on that, in all cases but
one, the functions $C^I$ must in fact be constants (and (\ref{vuit}) then
implies that $X_1$ is just a vector field on $M_1$). Thus, we conclude that
whenever $\Psi _{,\alpha }=0$, one has 
\begin{equation}
X=X_1^A(x^B,x^\gamma )\partial _A+C^I(x^B)\zeta _I  \label{homot}
\end{equation}
where $C^I$ and $X_1^A$ are functions of their arguments, to be determined
from (\ref{set}) and (\ref{vuit}), and $\{\zeta _I\}$ with $I\leq 4$ form a
basis of the homothetic algebra of $({\tilde{M}}_2,h_2)$.

Let us next study what happens on $H$. Notice that (\ref{x2conf4}) can be
rewritten as $\pounds _Yh_{\alpha \beta }=2\lambda h_{\alpha \beta }$ with $%
Y_\alpha =\Psi _{,\alpha }$; thus, $Y$ is also a CKV of $({\tilde{M}}_2,h_2)$
with conformal factor $\lambda $, and one therefore has \cite{Hall90}: 
\begin{equation}
\pounds _YR_{2\alpha \beta }=-2\lambda _{\alpha /\beta }-\left( h^{\mu \nu
}\lambda _{\mu /\nu }\right) h_{\alpha \beta }  \label{lambdaH}
\end{equation}
which, on account of (\ref{Escricci2}) and (\ref{x2conf4}), can be rewritten
as 
\begin{equation}
\lambda _{\alpha /\beta }=-\frac 18\left( R_{2,\alpha }\Psi ^\alpha
+2\lambda R_2\right) h_{\alpha \beta }\equiv \Sigma h_{\alpha \beta }
\label{asterisc}
\end{equation}
that is: $Z$ such that $Z_\alpha \equiv \lambda _{,\alpha }$ is a (gradient)
CKV, colinear with another CKV, namely $Y$; it is then immediate to show,
taking into account (\ref{x2conf4}), (\ref{lambdaderiv}), (\ref{R2deriv})
and (\ref{asterisc}), that $R_2$ must be constant ($\sigma =0$ in (\ref
{R2deriv})), (\ref{x2conf4}) then reading 
\begin{equation}
\Psi _{\alpha /\beta }=-\frac{R_2}4\Psi h_{\alpha /\beta }
\end{equation}
The Bianchi identities specialized to $\Psi _{,\alpha }$ then imply one of
the following:

\begin{enumerate}
\item  $R_2=0$ and $\Psi _{,\alpha }\not{=}0\ $, one then has the expression
(\ref{sckvK}) for $X_2$, etc.

\item  $R_2=$ constant ($\not{=}0$) and $\Psi _{,\alpha }=0\ $, $X_2$ is
then an HVF of $\left( {\tilde{M}}_2,h_2\right) \ $, but since $R_2$ is
constant and non-zero, it must be a KV, i.e.; $\Psi =0$.

\item  $R_2=0$ and $\Psi _{,\alpha }=0\ $, $X_2$ is an HVF of $\left( {%
\tilde{M}}_2,h_2\right) \ $, possibly non-Killing.
\end{enumerate}

Notice that whenever $\Psi_{,\alpha} = 0 \ $, one gets the same results as
when studying this case in $K \subset {\tilde M}_2$, i.e.; equations (\ref
{psinova}) and (\ref{homot}) hold.

We can roughly summarize the results so far obtained as follows:

The horizontal component $X_2$ of a RC $X$ is either an HVF of $({\tilde M}%
_2,h_2)$ (i.e.; $\Psi_{,\alpha}=0$) and $X$ is then given by (\ref{homot}),
or else it is a proper SCKV of $({\tilde M}_2,h_2)$ (that is; $%
\Psi_{,\alpha} \not{= }0$, $\Psi_{\alpha / \beta}=0$), this being possible
only when $R_2=0$ (i.e.; $({\tilde M}_2,h_2)$ flat), and in that case $X$
takes the form given by (\ref{sckvK}). In both cases, the functions
appearing in (\ref{homot}) and (\ref{sckvK}) must satisfy (\ref{set}) and (%
\ref{vuit}).

We shall next focus our attention on the case $X_2$ homothetic, studying the
various cases that may arise in connection with the different structures and
dimensions of the homothetic algebra of $({\tilde M}_2, h_2)$.

To this end, let ${\cal H}_r$ be the homothetic algebra of $({\tilde M}_2,
h_2)$, $r$ being its dimension. Since dim $M_2=2$ it follows that $r$ can
only be $0, \, 1, \, 2, \, 3$ or $4$. We shall deal separately with all
these cases assuming, for the sake of simplicity, that $h_2$ is Riemannian
(similar conclusions hold if $h_2$ is Lorentz).

\begin{enumerate}
\item  $r=0$. In this case no HVFs exist (including KVs), and therefore $%
\Psi =C^I=0$, i.e.; $X_2=0$ and $X=X_1$ with $X_{1,\alpha }^D=0$ as a
consequence of (\ref{vuit}), that is: $X$ is a vector field on $M_1$ which
must satisfy (\ref{set}) and (\ref{psinova}) with $\Psi =0$.

\item  $r=1$. There are now two cases to be distinguished, depending on
whether a proper HVF exists or not.

\begin{enumerate}
\item  A proper HVF $\zeta $, exists in $({\tilde{M}}_2,h_2)$. It is easy to
see that one can then always choose coordinates, say $x$ and $y$, such that
the line element $d\sigma ^2$ associated with $h_2$, and the HVF $\zeta $
read in these coordinates 
\begin{equation}
d{\sigma }^2=e^{2y}(dx^2+h^2(x)dy^2)\,\,\,{\rm and}\,\,\zeta =\partial _y
\end{equation}
the associated Ricci scalar is $R_2=-2e^{-2y}h^{-1}h^{\prime \prime }$ (a
prime denoting derivative with respect to $x$), and (\ref{vuit}) then
implies:

\begin{equation}
R_{AD}X_{1,x}^D=0\,\,\,\,\,\,{\rm and}\,\,\,\,\,\,R_{AD}X_{1,y}^D=-F\Psi
_{,A}e^{2y}h^2(x)
\end{equation}
which cannot be fulfilled unless $(Fh^2(x))_{,x}=0$, i.e.; $h(x)={\rm %
constant}$, in which case $R_2=0$ and therefore $r=4$. Thus, $\Psi _{,A}=0$ (%
$\Psi =$ constant $\not{=}0$), $X_{1,\alpha }^D=0$ and then $%
X=X_1^A(x^D)\partial _A+\Psi \zeta $ with $X_1$ satisfying (\ref{set}) and (%
\ref{psinova}) with $\Psi =$ constant ($\not{=}0$).

\item  No proper HVF exists in $({\tilde{M}}_2,h_2)$, just a KV, say $\xi $.
It then follows that $\Psi =0$ necessarily, and again coordinates may be
chosen such that 
\begin{equation}
d{\sigma }^2=dx^2+h^2(x)dy^2\,\,\,\,\,\,{\rm and}\,\,\,\,\,\xi =\partial _y
\end{equation}
the Ricci scalar is then $R_2=-2h^{-1}h^{\prime \prime }$, and (\ref{vuit})
implies, as in the previous case, $(Fh^2(x))_{,x}=0$, which in turn can be
seen to imply

\begin{equation}
\left( \Phi ^2\right) _{;A}^A=2a\,\,\,\,,\,\,\,\,(a={\rm constant})
\end{equation}
\begin{equation}
h^{\prime \prime }+ah^2=b\,\,\,\,,\,\,\,\,(b={\rm constant})
\end{equation}
Performing now the coordinate change $h(x)\equiv z$, the above line element
reads 
\begin{equation}
d{\sigma }^2={\frac{dz^2}{2C+z^2+2\log z}}+z^2dy^2
\end{equation}
Hence (\ref{vuit}) implies $X_1^A=P^A(x^D)y+Q^A(x^D)$ and then 
\begin{equation}
X=\left( P^A(x^D)y+Q^A(x^D)\right) \partial _A+C(x^D)\xi
\label{excepcional1}
\end{equation}
where $P^A(x^D)$ and $Q^A(x^D)$ must both satisfy (\ref{set}) separately,
and $C(x^D)$ must be such that $R_{AD}P^D=-bC_{,A}$.
\end{enumerate}

\item  $r=2$, ${\cal H}$ must then contain at least one proper HVF, since
otherwise (${\cal H}$ spanned by two KVs) a third KV would necessarily
exist, hence dim ${\cal H}=3$. Suppose then that a proper HVF, $\zeta $,
exists; the other vector in the basis of ${\cal H}$ can always be chosen to
be a KV, say $\xi $, and there are two possible, non-isomorphic, Lie algebra
structures for ${\cal H}$, namely $[\xi ,\zeta ]=0$ (abelian), and $[\xi
,\zeta ]=\xi $ (non abelian). In the abelian case, coordinates may be chosen
such that the line element, $\zeta $ and $\xi $ read respectively 
\begin{equation}
d\sigma ^2=dx^2+x^2dy^2\,\,,\,\,\zeta =x\partial _x\,\,,\,\,\xi =\partial _y
\end{equation}
but it then follows that $R_2=0$ and therefore two other KVs exist, $r$ thus
being 4, therefore this case cannot arise.

In the non-abelian case, and again by means of a suitable choice of
coordinates, one has: 
\begin{equation}
d\sigma ^2=dx^2+x^{2{\frac{n-1}n}}dy^2\,\,,\,\,\zeta =nx\partial
_x+y\partial _y\,(n\not{=}1)\,\,,\,\,\xi =\partial _y
\end{equation}
but then (\ref{vuit}) implies, as in previous cases, that $\left( Fx^{2{%
\frac{n-1}n}}\right) _{,x}=0$, which can not be satisfied, therefore $\Psi
_{,A}=C_{,A}=0$ (i.e.; $\Psi $ and $C$ constants) and then $X_{1,\alpha
}^D=0 $, and again $X=X_1^A(x^D)\partial _A+\Psi \zeta $ with $X_1$
satisfying (\ref{set}) and (\ref{psinova}) with $\Psi =$ constant ($\not{=}0$%
).

\item  $r=3$ If a proper HVF exists in $({\tilde{M}}_2,h_2)$, the associated
Killing subalgebra is then of dimension 2, and therefore a third KV exists,
hence dim ${\cal H}=4$ and therefore this case is not possible. If, on the
other hand, no proper HVFs exist, $({\tilde{M}}_2,h_2)$ is of constant
curvature and $\Psi =0$ necessarily. Let $\{\xi _J\}\,,\,\,J=1,2,3$ be three
KVs spanning ${\cal H}$, from (\ref{vuit}) it follows $R_{AD}X_{1,\alpha
}^D=-FC_{,A}^J\xi _{J\alpha }$, differentiating with respect to $x^\beta $,
skewsymmetrising and equating to zero, one has 
\begin{equation}
C_{,A}^J\xi _{J[\alpha ,\beta ]}=0
\end{equation}
that is; either $C_{,A}^J=0$ or else $({\tilde{M}}_2,h_2)$ contains a
gradient KV. From \cite{KramerEtal80}, it is easy to see that the later is
only possible if $R_2=0$, but in that case a proper HVF is always admitted
(namely $\zeta =x\partial _x$ in the coordinates used in \cite{KramerEtal80}%
), and therefore dim ${\cal H}=4$.

\item  $r=4$. In this case $({\tilde{M}}_2,h_2)$ is flat, the line element
and KVs being those given in \cite{KramerEtal80} and the proper HVF $\zeta
=x\partial _x$. Proceeding as before, one can readily see from (\ref{vuit})
that $X_1^A=M^A(x^D)x^2+(P_1^A(x^D)\cos y+P_3^A(x^D)\sin y)x+Q^A(x^D)$, but
since $\Psi \not{=}0$ and $\Psi _{,\alpha }=0$, it follows that $%
P_1^A=P_3^A=M^A=0$, which in turn imply $\Psi
_{,A}=C_{,A}^1=C_{,A}^2=C_{,A}^3=0$, hence $X_{1,\alpha }^A=0$, that is: $%
X_1 $ is a vector field on $M_1$ that has to satisfy (\ref{set}) and (\ref
{psinova}) with $\Psi =$ constant ($\not{=}0$), and $X=X_1^A(x^D)\partial
_A+\Psi \zeta +C^J\xi _J$.
\end{enumerate}

Our purpose in the next sections is to apply the results so far obtained to
the case of spherically symmetric space-times which are also static, as well
as to Friedmann-Robertson-Walker (FRW) models.

\section{Spherically symmetric space-times}

We next specify the above results to the case of a general spherically
symmetric spacetime whose metric, in the local chart $\{x^{0,1,2,3}~= ~t,r,
\vartheta ,\phi \}$ takes the form \cite{KramerEtal80} 
\begin{equation}  \label{MetricShearF}
ds^2=\ -{\bf e}^{2\nu (t,r)}\,{\rm d}t^2+\,{\bf e}^{2\lambda (t,r)}\, {\rm d}
r^2+\,r^2({\rm d}\vartheta ^2+\,\sin ^2\phi {\rm \,d}\phi ^2)
\end{equation}
Comparing the metric (\ref{warped}) with the above (\ref{MetricShearF}), we
have $\{x^A~=~t,\ r\ ;\ x^\alpha =\vartheta ,\ \phi \}$ ; $\Phi ~=~r$ ;

\[
h_{AB}\left( t,r\right) {\rm d}x^A{\rm d}x^B~=~\,-{\bf e}^ {2\nu (t,r)} {\rm %
d}t^2~~+~~\,{\bf e}^{2\lambda (t,r)} \,{\rm d}r^2 
\]

and

\[
h_{\alpha \beta }{\rm d}x^\alpha {\rm d}x^\beta ~=~{\rm d} \vartheta
^2~+~\sin ^2\phi {\rm \,d}\phi ^2\ 
\]
Thus, the Ricci tensor can be written as

\begin{equation}  \label{sphricci1}
R_{tt}=-\frac 12R_1{\bf e}^{2\nu (t,r)}\ +\frac{2\nu ^{\prime }}r\ {\bf e}
^{2\left( \nu (t,r)-\lambda (t,r)\right) }
\end{equation}
\begin{equation}  \label{sphriccitr}
R_{t\ r}=\frac 2r\ \dot{\lambda}
\end{equation}
\begin{equation}  \label{sphricci2}
R_{rr}=\frac 12\ R_1\ {\bf e}^{2\lambda }+\frac{2\lambda ^{\prime }}r
\end{equation}
and 
\begin{equation}  \label{sphricci3}
R_{\alpha \beta }=\left\{ 1-{\bf e}^{-2\lambda }\left[ 1+r\ \left( \nu
^{\prime }-\lambda ^{\prime }\right) \right] \right\} h_{\alpha \beta }
\end{equation}
where a dash and a dot indicate, as usual, partial derivatives with respect
to $r$ and $t$ respectively. As above, $R_1$ is the Ricci scalar associated
with the 2-dimensional metric $h_{AB}$, and now $\frac 12~R_2~=~1$.

According to our foregoing discussion, any RC $X$ must be of the form 
\begin{equation}  \label{rc}
X=X_1+C^J \xi _J
\end{equation}
where $\left\{ \xi _J,\ \ J=1,2,3\right\} $ are the KV 's that implement the
spherical symmetry, $X_1=X^A(t,r)\partial _A$ and $C^J$ are constants, $%
J=1,\ 2, \ 3$, which can be set equal to zero without loss of generality
(since $C^J \xi_J$ is a KV of the space-time and therefore a trivial RC). On
the other hand, since $\Psi=0$ and $(\Phi^2)^A_{;A}= 2{\bf e}^{-2\lambda }
\left[ 1+r\ \left( \nu ^{\prime }-\lambda ^{\prime }\right) \right]$, (\ref
{psinova}) implies 
\begin{equation}  \label{psi1}
\{{\bf e}^{-2\lambda } \left[ 1+r\ \left( \nu ^{\prime }-\lambda ^{\prime
}\right) \right] \}_{,D} X^D =0
\end{equation}
therefore, the proper RCs of a spherically symmetric space-time whose Ricci
tensor is non-degenerate, are of the form 
\begin{equation}
X=X^t(t,r) \partial_t + X^r(t,r) \partial_r
\end{equation}
and they must satisfy (\ref{psi1}) in addition to (\ref{set}) specialised to
the Ricci tensor components given by (\ref{sphricci1}), (\ref{sphriccitr})
and (\ref{sphricci2}).

We shall next present two examples: static spherically symmetric
space-times, and FRW spacetimes. 

\subsection{Static spherically symmetric space-times}

Let us consider first the case of static spherically symmetric spacetimes,
these are described by (\ref{MetricShearF}) where the functions $v$ and $%
\lambda $ appearing in it depend just on $r$ , $\partial _t$ thus being a
KV. For the purpose of this paper it is convenient to write the components
of the Ricci tensor for this metric as follows \cite{BokhariQadir93}, \cite
{JamilEtal94}, \cite{FaridEtal95} 
\begin{equation}
R_{tt}\equiv A(r)\quad R_{rr}\equiv B(r)\quad R_{\theta \theta }\equiv
C(r)\quad {\rm and}\quad R_{\phi \phi }\equiv \sin ^2\theta \ R_{\theta
\theta }  \label{Ricciform}
\end{equation}

Taking now into account the results of the previous section one has 
\begin{equation}  \label{GeneralColl}
X =X ^t\left( t,r\right) \partial _t+X ^r\left( t,r \right) \partial _r
\end{equation}
and the (non-trivial) equations arising from (\ref{set}), are simply: 
\begin{equation}  \label{LieRicci00}
A^{\prime }(r)X ^r+2A(r)X _{,t}^t=0
\end{equation}
\begin{equation}  \label{LieRicci01}
A(r)X _{,r}^t+B(r)X _{,t}^r=0
\end{equation}
\begin{equation}  \label{LieRicci11}
B^{\prime }(r)X ^r+2B(r)\ X _{,r}^r=0
\end{equation}
\begin{equation}  \label{LieRicci22}
C^{\prime }(r)X ^r=0
\end{equation}
Equation (\ref{LieRicci22}) directly implies: 
\begin{equation}  \label{c'=0}
C^{\prime }(r)=0
\end{equation}
since otherwise one would have $X ^r=0$ that would imply, from the remaining
equations, $X ^t=$ constant, thus being a KV and not a proper RC.

A direct integration of equation (\ref{LieRicci11}) gives 
\begin{equation}  \label{tempXi1}
X ^r=\frac{{\cal K}(t)}{\sqrt{\left| B(r)\right| }}
\end{equation}
Now, substituting this result back into eq. (\ref{LieRicci00}) and (\ref
{LieRicci01}), differentiating them with respect to $t$ and $r$,
respectively; and equating the crossed derivatives of $X ^t$, we obtain 
\begin{equation}  \label{temp11Xi}
{\cal K}_{,tt}\ \frac{\sqrt{\left| B(r)\right| }}{A(r)}=\frac 12 {\cal K}
\left( \frac{A^{\prime }(r)\ }{A(r)\sqrt{\left| B(r)\right| }} \right)
^{\prime }
\end{equation}
and the following two cases arise:

\subsection{\bf Case I}

\begin{equation}  \label{temp12Xi}
{\cal K}_{,tt}\ -\epsilon \ k^2{\cal K}=0;\qquad k={\it Const},\quad
\epsilon =\pm 1
\end{equation}
therefore 
\begin{equation}  \label{temp13Xi}
{\cal K}(t)=\left\{ 
\begin{array}{cc}
a{\bf e}^{kt}+b{\bf e}^{-kt} & \epsilon =+1 \\ 
a\sin kt+b\cos kt & \epsilon =-1
\end{array}
\right| \quad
\end{equation}
and 
\begin{equation}  \label{temp14Xi}
2\epsilon \ k^2\ \frac{\sqrt{\left| B(r)\right| }} {A(r)}=\left( \frac{
A^{\prime }(r)\ }{A(r)\sqrt{\left| B(r)\right| }} \right) ^{\prime }
\end{equation}
Substituting these results back into (\ref{LieRicci00}), integrating and
plugging them back into (\ref{LieRicci01}), we find 
\begin{equation}  \label{temp15Xi}
X ^t=-\frac 12 \left( \frac{A^{\prime }(r)\ }{A(r)\sqrt{\left| B(r) \right| }
}\right) M(t)
\end{equation}
where $M(t)= \int {\cal K}(t) dt$. and the constant of integration has been
set equal to zero without loss of generality.

Thus, for this case a proper RC is of the form:

\begin{equation}  \label{Xisolcas1}
X =-\frac 12\left( \frac{A^{\prime }(r)\ }{A(r)\sqrt{\left| B(r) \right| }}
\right) \left( \int {\cal K}(t){\rm d}t\right) \partial _t+ \frac{{\cal K}%
(t) }{\sqrt{\left| B(r)\right| }}\partial _r
\end{equation}
where ${\cal K}(t)$ is given by (\ref{temp13Xi}), and the components of the
Ricci tensor must satisfy (\ref{c'=0}) and (\ref{temp14Xi}).

\subsection{\bf Case II}

\begin{equation}  \label{II-1}
{\cal K}={\cal S}_1t+{\cal S}_2\qquad {\cal S}_1, \ {\cal S}_2 = {\it Const}
\end{equation}
and 
\begin{equation}  \label{II-2}
\frac 12\frac{A^{\prime }(r)\ }{A(r)\sqrt{\left| B(r)\right| }}= \sigma =%
{\it Const}
\end{equation}
Then from (\ref{tempXi1}) and (\ref{LieRicci00}-\ref{LieRicci11}), one gets
after some straightforward calculations:

\begin{equation}  \label{temp30Xi}
X =\left\{ -\sigma \left( \frac 12{\cal S}_1\ t^2+{\cal S}_2 \ t \right) + 
\frac{{\cal S}_1}{2\,\sigma }\frac 1{A(r)}\right\} \, \partial _t+\left( 
{\cal S}_1\ t+{\cal S}_2\ \right) \frac 1 {\sqrt{B(r)}}\,\partial _r
\end{equation}

As an example of a space-time satisfying the above requirements \cite
{BertolotiEtal95b}, take for instance 
\begin{equation}
\nu (r)=\frac 12\left( \frac{r^4}{8r_0^2}+h\,\ln \frac r{r_0}+k\right) \quad 
{\rm and\quad }\lambda (r)=\nu (r)+\,\ln \frac r{r_0}  \label{metricelem}
\end{equation}
therefore the Ricci components can be written 
\begin{equation}
B(r)=2\frac{h+1}{r^2}\qquad C(r)=1\quad {\rm and}\quad A(r)={\em Const}
\label{b}
\end{equation}
where $r_0$, $h$ and $k$ are constants, and we have for the RC: 
\begin{equation}
X^t=-\,c_4\sqrt{2\,\left( h+1\right) }\,\ln r+c_0\quad {\rm and}\quad X^r=%
\frac{c_4\,t+c_5}{\sqrt{2\,\left( h+1\right) }}r  \label{Collstatic}
\end{equation}
This result invalidates a misleading theorem stated in reference \cite
{JamilEtal94}, and used in \cite{FaridEtal95}. According to this
``theorem'', this collineation vector (\ref{Collstatic}) should represent an
isometry; however it is easy to see that $X$ does not reduce to a KV unless $%
c_4~=~c_5~=~0$.

Since all KV's are naturally RC's and these (if assumed smooth) form a Lie
algebra under the usual bracket operation, the Lie bracket of the above RC's
with the four KV's the metric admits, must yield in turn RC's; thus 
\[
\left[ \xi _I,\xi \right] =0\qquad \forall I=1,2,3
\]
where $\xi _I$ designate the KV's implementing the spherical symmetry, and 
\[
\left[ \partial _t,X\right] =X^{\prime }\left( \neq 0\right) 
\]
where $X^{\prime }=\frac{c_{4\,}r}{\sqrt{2(h+1)}}$ is also a proper RC.

A more detailed account of RCs for non-static spherically symmetric
space-times will be given in a forthcoming paper.

\subsection{FRW space-times.}

As an example of RC for non-static spherically symmetric metrics, we
consider FRW space-times described by \cite{Weinberg72}:

\begin{equation}  \label{frw}
{\rm d}s^2=-\ {\rm d}t^2+R\left( t\right) ^2\left( \frac{{\rm d}r^2}
{1-k\,r^2 }+r^2\,{\rm d}\vartheta ^2+r^2\sin ^2\vartheta \ {\rm d} \phi
^2\right)
\end{equation}

Again, using the above notation, we have $\Phi =rR(t)$, 
\begin{equation}  \label{subspace1}
h_{AB}\,{\rm d}x^A\,{\rm d}x^B=-\ {\rm d}t^2+R\left( t\right) ^2 \frac{{\rm d%
} r^2}{1-k\,r^2}
\end{equation}
and 
\begin{equation}  \label{subspace2}
h_{\alpha \beta }\,{\rm d}x^\alpha \,{\rm d}x^\beta =\, {\rm d}\vartheta
^2+\sin ^2\vartheta \ {\rm d}\phi ^2
\end{equation}
Then the Ricci tensor takes the form 
\begin{eqnarray}
&&R_{tt}=-3\frac{\ddot{R}}R  \nonumber \\
&&R_{rr}=g_{rr}\frac \Delta {R^2}  \label{ricci-frw} \\
&&R_{\alpha \beta }=g_{\alpha \beta } \frac \Delta {R^2}  \nonumber \\
&&\Delta =2k+2\dot{R}^2+R\ddot{R}  \nonumber
\end{eqnarray}
Specializing (\ref{set}) to the present case, we obtain 
\begin{eqnarray}
&&X ^tR_{rr,t}+X ^rR_{rr,r}+2R_{rr}X _{,r}^r=0  \nonumber \\
&&X ^tR_{tt,t}+2R_{tt}X _{,t}^t=0  \label{rc-frw} \\
&&R_{tt}X _{,r}^t+R_{rr}X _{,t}^r=0  \nonumber \\
&&X ^tR_{\theta \theta ,t}+X ^rR_{\theta \theta ,r}=0  \nonumber
\end{eqnarray}
Thus, we get \cite{NunezEtal90} 
\begin{eqnarray}
X ^t &=&c\left( 1-kr^2\right) ^{1/2}\left| R_{00}\right| ^{-1/2}  \nonumber
\\
X ^r &=&g(t)\,r\,\,\left( 1-kr^2\right) ^{1/2}  \label{psi-frw}
\end{eqnarray}
where $g(t)=-\,c\,\left| R_{00}\right| ^{-1/2}\left( \dot \Delta /2\Delta
\right) $ , and $c$ is a constant.

\section{ACKNOWLEDGEMENTS}

Two of us (J.C. and U. P.) gratefully acknowledge funding from Postgrado en
Astronom\'{\i}a y Astrof\'{\i}sica as well as the warm hospitality of the
Laboratorio de F\'{\i}sica Te\'{o}rica, Universidad de Los Andes M\'{e}rida,
Venezuela, where most of this work was carried out. J.C. acknowledges
partial financial support from STRIDE program, Research Project No.
STRDB/C/CEN/509/92. The authors wish also to thank the staff of the {\it SUMA%
}{\em \ }, the computational facility of the Faculty of Science (Universidad
de Los Andes).

\end{document}